\documentclass[twocolumn,amssymb,amsmath,nobibnotes,
  preprintnumbers,showpacs,prl]{revtex4}
\usepackage{epsfig,scalefnt}

\newcommand{\abbrev}{\scalefont{.9}}
\newcommand{\drbar}{$\overline{\mbox{\abbrev DR}}$}

\newcommand{\lhc}{{\abbrev LHC}}
\newcommand{\qcd}{{\abbrev QCD}}
\newcommand{\mssm}{{\abbrev MSSM}}
\newcommand{\susy}{{\abbrev SUSY}}

\newcommand{\dred}{{\abbrev DRED}}

\newcommand{\mtop}{M_t}

\newcommand{\Mstop}[1]{M_{\tilde t_{#1}}}

\newcommand{\Mgluino}{M_{\tilde g}}
\newcommand{\muSUSY}{\mu_{\rm \susy{}}}
\newcommand{\lmMtMS}{L_{tS}}

\newcommand{\lmMtMsq}{L_{t\tilde{q}}}
\newcommand{\lmuMt}{L_{{\mu}t}}

\begin{document}

\preprint{SFB/CPP-08-14, TTP08-08, WUB08-02}


\title{
  Higgs boson mass in supersymmetry to three loops
}
\author{R.V. Harlander$^{1}$, P. Kant$^{2}$,
  L. Mihaila$^{2}$, M. Steinhauser$^{2}$}
\affiliation{$^1$ Fachbereich C, Theoretische Physik,
  Universit{\"a}t Wuppertal,
  42097 Wuppertal, Germany\\
  $^2$ Institut f{\"u}r Theoretische Teilchenphysik,
  Universit{\"a}t Karlsruhe,
  76128 Karlsruhe, Germany}

\date{\today}

\begin{abstract}
  Within the minimal supersymmetric extension of the Standard Model, the
  mass of the light CP-even Higgs boson is computed to three-loop
  accuracy, taking into account the next-to-next-to-leading order
  effects from supersymmetric Quantum Chromodynamics.  We consider two
  different scenarios for the mass hierarchies of the supersymmetric
  spectrum. Our numerical results amount to corrections of about
  500~MeV which is of the same order as the experimental
  accuracy expected at the CERN Large Hadron Collider (LHC).
\end{abstract}
\pacs{11.30.Pb, 12.38.-t, 14.80.Cp}

\maketitle



\section{I. Introduction}


Supersymmetry is currently the most-studied extension of the Standard
Model~(see, e.g., Ref.\,\cite{Nilles:1983ge}). It provides solutions to
some profound theoretical problems of the Standard Model: the fine
tuning of the Higgs mass, the (non-)unification of gauge couplings, a
mechanism for spontaneous symmetry breaking, and a Cold Dark Matter
candidate.

The minimal supersymmetric extension of the Standard Model (\mssm{}) is
based on a two-Higgs-doublet model ({\abbrev 2HDM}) with five physical
Higgs bosons: two {\abbrev CP}-even $h/H$, one {\abbrev CP}-odd $A$
(also named the ``pseudo-scalar'' Higgs), and two charged scalars
$H^\pm$. Each particle of this {\abbrev 2HDM} receives a \susy{} partner
of opposite spin-statistics, where left- and right-handed components of
a Standard Model Dirac fermion are attributed with separate scalars
$\tilde f_{\rm L/R}$ which mix to the physical mass eigenstates $\tilde
f_{1/2}$.

Compared to the Standard Model, the \mssm{} Higgs sector is described by
two additional parameters, usually chosen to be the pseudo-scalar mass
$M_A$ and the ratio of the vacuum expectation values of the two Higgs
doublets, $\tan\beta=v_2/v_1$. The masses of the other Higgs bosons are
then fixed by \susy{} constraints. In particular, the mass of the light
{\abbrev CP}-even Higgs boson, $M_h$, is bounded from above. At
tree-level, it is $M_h<M_Z$. Radiative corrections to the Higgs pole
masses raise this bound substantially to values that were inaccessible
by {\abbrev LEP}~\cite{Ellis:1990nz,Okada:1990vk,Haber:1990aw}. The
large numerical impact is due to a contribution $\sim \alpha_t M_t^2\sim
M_t^4$ coming from top- and stop quark loops ($M_t$ is the top quark
mass and $\sqrt{\alpha_t}$ is proportional to the top Yukawa coupling).

The one-loop corrections to the Higgs pole masses are known without any
approximations~\cite{Chankowski:1991md,Brignole:1992uf,
Dabelstein:1994hb,Pierce:1996zz}.  They show that the bulk of the
numerical effects can be obtained in the so-called effective-potential
approach in the limit of vanishing external momentum. Motivated by this
observation, all presumably relevant two-loop terms have since been
evaluated in this approach (for reviews, see e.g.\
Refs.\,\cite{Heinemeyer:2004ms,Allanach:2004rh}).  More recently there
has been quite some activity in the context of the \mssm{} with complex
parameters which can lead to sizeable effects (see, e.g.,
Ref.~\cite{Heinemeyer:2007aq}).  The two-loop results are implemented in
the numerical programs {\sf FeynHiggs}~\cite{FeynHiggs} and {\sf
CPsuperH}~\cite{Lee:2003nta,Lee:2007gn} using on-shell particle masses,
and in {\sf SoftSusy}~\cite{Allanach:2001kg}, {\sf
SPheno}~\cite{Porod:2003um}, and {\sf Suspect}~\cite{Djouadi:2002ze}
using $\overline{\rm DR}$ parameters, that is, dimensional reduction
with minimal subtraction.  The influence of terms that go beyond the
approximation of vanishing external momentum has been investigated in
Ref.\,\cite{Martin:2001vx:2002iu}.

Based mostly on the renormalization scale and scheme dependence, the
theoretical uncertainty on the prediction of the light Higgs boson mass
$M_h$ has been estimated to
3-5\,GeV~\cite{Degrassi:2002fi,Allanach:2004rh}. This is to be compared
with the expected experimental uncertainty of a Higgs mass measurement
at the \lhc{} of the order of 100-200~MeV~\cite{cmstdr}. At an
International Linear Collider, this goes even down to roughly
50~MeV~\cite{AguilarSaavedra:2001rg}. These numbers clearly show the
need for three-loop corrections to the \susy{} Higgs bosons masses in
order to fully exploit the physics potential of these colliders.

In fact, quite recently the leading and next-to-leading 
logarithmic terms in $\ln(M_{\rm \susy{}}/M_t)$
at three-loop level have been obtained, where $M_{\rm
  \susy{}}$ is the typical scale of \susy{} particle
masses~\cite{Martin:2007pg}. In this letter, we want to present the
first genuine three-loop calculation of the lightest Higgs boson mass,
focusing on a few simplifying limiting cases for the sake of brevity.
In particular, we consider effects of order $\alpha_t\alpha_s^2$, keep
only the leading terms $\sim M_t^4$, and neglect all mixing effects in
the stop sector.  More general results and their detailed
phenomenological impacts shall be deferred to a later publication.


\section{II. The Higgs boson mass in the MSSM}


At tree-level, the mass matrix of the neutral, {\abbrev CP}-even Higgs
bosons $h$, $H$ has the following form:
\begin{eqnarray}
  \lefteqn{
  {\cal M}_{H,\rm tree}^2 =
  \frac{\sin 2\beta}{2}\times}
\\&&
  \left(
  \begin{array}{cc}
    M_Z^2 \cot\beta + M_A^2 \tan\beta &
    -M_Z^2-M_A^2 \\
    -M_Z^2-M_A^2 &
    M_Z^2 \tan\beta + M_A^2 \cot\beta
  \end{array}
  \right)
  \,.
\nonumber
\end{eqnarray}
The diagonalization of ${\cal M}_{H,\rm tree}^2$ gives the tree-level result
for $M_h$ and $M_H$, and leads to the well-known bound $M_h < M_Z$ which
is approached in the limit $\tan\beta\to \infty$.

Quantum corrections to the Higgs boson masses are incorporated by
evaluating the poles of the Higgs boson propagator at higher orders. As
mentioned in the Introduction, the numerically dominant contributions
can be obtained in the approximation of zero external momentum (see,
e.g., Refs.~\cite{Heinemeyer:1998jw:1998kz:1998np:1999be}) which we will
adopt in the following.  Furthermore, we will only consider corrections
of order $\alpha_t\alpha_s^2$.  Apart from the quark, squark, and gluino
masses, there is another parameter with mass dimension, the trilinear
coupling of the soft \susy{} breaking terms, $A_t$.  Before
renormalization, we express it through the stop masses $\Mstop{1}$,
$\Mstop{2}$, the stop mixing angle $\theta_t$, and the bilinear Higgs
parameter $\muSUSY$ as follows:
\begin{eqnarray}
  2 \mtop A_t &=& (\Mstop{1}^2-\Mstop{2}^2)\sin 2\theta_t 
  + 2 \mtop \muSUSY \cot\beta
  \,.
  \label{eq::Atmu}
\end{eqnarray}

The mass matrix ${\cal M}_H^2$ is obtained from the quadratic terms in
the Higgs boson potential constructed from the fields $\phi_1$ and
$\phi_2$. They are related to the physical Higgs mass eigenstates via a
mixing angle $\alpha$.  Since $\phi_1$ does not couple directly to top
quarks, it is convenient to perform the calculations of the Feynman
diagrams in the $(\phi_1,\phi_2)$ basis.

Including higher order corrections, one obtains the Higgs boson mass
matrix
\begin{eqnarray}
  {\cal M}_{H}^2 &=&
  {\cal M}_{H,\rm tree}^2 -
  \left(
  \begin{array}{cc}
    \hat\Sigma_{\phi_1}       & \hat\Sigma_{\phi_1\phi_2} \\
    \hat\Sigma_{\phi_1\phi_2} & \hat\Sigma_{\phi_2}
  \end{array}
  \right)
  \,,
  \label{eq::MH}
\end{eqnarray}
which again gives the physical Higgs boson masses upon diagonalization.
The renormalized quantities $\hat\Sigma_{\phi_1}$, $\hat\Sigma_{\phi_2}$
and $\hat\Sigma_{\phi_1\phi_2}$ are obtained from the self energies of
the fields $\phi_1$, $\phi_2$, $A$, evaluated at zero external
momentum, as well as from tadpole contributions of $\phi_1$ and $\phi_2$
(see, e.g., Ref.\,\cite{Heinemeyer:2004ms}).  Let us remark that if one
sets $\Mstop{1}=\Mstop{2}$ and $A_t=0$, and evaluates only the leading
contribution $\sim\mtop^4$, then only $\hat\Sigma_{\phi_2}\neq 0$ and
the matrix ${\cal M}_{H}^2-{\cal M}_{H,\rm tree}^2$ is diagonal. On the
other hand, if we allow for non-zero $A_t$, also $\hat\Sigma_{\phi_1}$
and $\hat\Sigma_{\phi_1\phi_2}$ contribute in general.


The calculation of $\hat\Sigma_{\phi_2}$ is organized as follows: 
All Feynman diagrams are generated with {\tt
  QGRAF}~\cite{Nogueira:1991ex}. In order to 
properly take into account the Majorana character of the gluino, the
output is subsequently manipulated by a {\tt PERL} script which
applies the rules given in Ref.~\cite{Denner:1992vza}. The various
diagram topologies are identified and transformed to {\tt
  FORM}~\cite{Vermaseren:2000nd} with the help of {\tt q2e} and {\tt
  exp}~\cite{Harlander:1997zb,Seidensticker:1999bb}.  The program {\tt
  exp} is also used in order to apply the asymptotic expansion (see,
e.g., Ref.~\cite{Smirnov:2002pj}) in the various mass hierarchies. The
actual evaluation of the integrals is performed with the package {\tt
  MATAD}~\cite{Steinhauser:2000ry}, resulting in an expansion in
$d-4$ for each diagram, where $d$ is the space-time dimension.  
The total number of three-loop diagrams
amounts to about 16,000.

At three-loop level we need to renormalize the top quark mass, the top
squark mass, and the stop mixing angle at the two-loop order. In
addition, the one-loop counterterm of the gluino mass is needed for the
renormalization of the two-loop expression.  We implement Dimensional
Reduction (\dred{}) with the help of the so-called $\epsilon$-scalars
which appear for the first time at two loops. The renormalization of the
$\epsilon$-scalar mass is performed in the on-shell scheme, requiring
that the renormalized mass is equal to zero. In the literature this is
referred to as \drbar{}$^\prime$ scheme.

The one-loop on-shell counterterms are well-known (see, e.g.,
Refs.~\cite{Pierce:1996zz,Djouadi:1998sq,Harlander:2004tp,Harlander:2005wm}).
As far as the two-loop counterterms for the squarks and quarks are
concerned, one can find the results in
Refs.~\cite{Martin:2005eg,Martin:2005ch}. However, it is rather tedious
to extract the results for the mass hierarchies we are interested in.
Thus, we re-computed the corresponding corrections.

To our knowledge, the two-loop counterterm for the stop mixing angle is
not yet available in the literature. It turns out that in our
approximation, where $\Mstop{1}=\Mstop{2}$ and $A_t=0$, 
only the one-loop
counterterm of the mixing angle enters the three-loop result.

As a cross check for our calculation, 
we recalculated the exact two-loop result (in the limit of
vanishing external momentum) and find perfect agreement with 
the literature~\cite{Heinemeyer:1998jw:1998kz:1998np:1999be,Degrassi:2001yf}.

Furthermore,
the expansion of the exact expressions confirms the limiting cases
discussed below.  Both the two- and three-loop calculations are
performed for a general \qcd{} gauge parameter $\xi_S$.  The
independence of the final results on $\xi_S$ serves as another welcome check
on the correctness of our result.

We use anti-commuting $\gamma_5$ which is allowed for fermion traces
which involve an even number of $\gamma_5$ matrices.  It turns out that all
traces involving an odd number of $\gamma_5$ vanish because they
contain less than four gamma matrices.

In the following we discuss three different cases for the mass
hierarchy.  In all cases we set the light quark masses to zero.


(i) {\it Supersymmetric limit}, i.e., $\mtop=\Mstop{}$ and the gluino
and other squarks are massless: $\Mgluino=M_{\tilde{q}}=0$.
The quantum corrections to the Higgs boson mass vanish in this case, as
required by supersymmetry.  Still, the individual diagrams are different
from zero and thus the calculation imposes a strong check on our setup.


(ii) {\it Massless gluino}, $\Mgluino =0$.  Expanding in the limit
$\mtop\ll \Mstop{}=M_{\tilde{q}} \equiv M_{\rm SUSY}$, we obtain for the
leading term of this expansion 
\begin{eqnarray}
  \hat\Sigma_{\phi_2} &=&
  \frac{3 G_F M_t^4}{\sqrt{2} \pi^2 \sin^2\beta}
  \Bigg\{\lmMtMS
  + \frac{\alpha_s}{\pi}\left[
    - 1 
    - 4\lmMtMS 
    + 2\lmMtMS^2
  \right]
  \nonumber\\&&\mbox{}
  + \left(\frac{\alpha_s}{\pi}\right)^2\left[
    - \frac{593}{27}
    - \frac{3}{4}\lmuMt
    + \frac{23}{81}\pi^2
    + \frac{401}{18}\zeta_3
    \right.\nonumber\\&&\left.\mbox{}
    + 
    \left(
      -\frac{47}{4}
      - 3\lmuMt 
      + \frac{4}{9}\pi^2
      - \frac{4}{9}\pi^2\ln2
    \right) \lmMtMS
      \right.\nonumber\\&&\left.\mbox{}
    + \left(
        \frac{1}{4}
      + \frac{3}{2}\lmuMt
    \right) \lmMtMS^2 
    + \frac{5}{2}\lmMtMS^3
    \right]
  \Bigg\}
  \,,
\end{eqnarray}
with $\lmuMt=\ln(\mu^2/M_t^2)$ and $\lmMtMS=\ln(M_t^2/M_{\rm SUSY}^2)$.


(iii) {\it Common \susy{} mass.}  In this scenario we assume
$\mtop\ll\Mstop{1}=\Mstop{2}=\Mgluino\equiv M_{\rm SUSY}\ll
M_{\tilde{q}}$.  Even though the top squark masses are equal and thus
the mixing angle is zero, it is necessary to introduce a counterterm for
$\theta_t$. Since the latter has contributions proportional to
$1/(\Mstop{2}^2-\Mstop{1}^2)$, we expand the one- and two-loop result in
this limit before inserting the counterterms. The
cancellation of such terms in the final result provides another check on
our calculation.

It is important to keep $A_t\neq 0$ in the one- and two-loop
contributions and to use Eq.~(\ref{eq::Atmu}) before renormalization,
because the corresponding counterterms generate terms of order $\mtop^4$
at three-loop level.  Setting $A_t=0$ in the end, we obtain 
\begin{widetext}
\begin{eqnarray}
  \hat\Sigma_{\phi_2} &=&
  \frac{3 G_F M_t^4}{\sqrt{2} \pi^2 \sin^2\beta}
  \Bigg\{\lmMtMS
  + \frac{\alpha_s}{\pi}\left[-4\lmMtMS + 2\lmMtMS^2\right]
  + \left(\frac{\alpha_s}{\pi}\right)^2\left[
      \frac{671}{324} 
    + \frac{1}{27}\pi^2
    + \frac{1}{9}\zeta_3
    + \left(
     - \frac{1591}{108} 
    - 3 \lmuMt
    + \frac{1}{3}\pi^2
    - \frac{4}{9}\pi^2\ln2
    \right.\right.\nonumber\\&&\left.\left.\mbox{}
    + \frac{55}{18}\lmMtMsq
    + \frac{5}{6}\lmMtMsq^2
    \right) \lmMtMS
    + \left(
      \frac{19}{18} 
     + \frac{3}{2}\lmuMt
     - \frac{5}{3}\lmMtMsq
    \right) \lmMtMS^2 
    + \frac{53}{18}\lmMtMS^3
    + \left( \frac{475}{108} - \frac{5}{9}\pi^2\right)\lmMtMsq 
    - \frac{25}{36}\lmMtMsq^2 
    - \frac{5}{18}\lmMtMsq^3
    \right.\nonumber\\&&\left.\mbox{}
    + {\cal O}\left(\frac{M_S^2}{M_{\tilde{q}}^2}\right)
    \right]
  \Bigg\}
  \,,
  \label{eq::sigphi2}
\end{eqnarray}
\end{widetext}
where 
$\lmMtMsq=\ln(M_t^2/M_{\tilde{q}}^2)$.  In Eq.~(\ref{eq::sigphi2}) we
only display the leading term in the $1/M_{\tilde{q}}$ expansion. We actually
computed five expansion terms and observe a rapid convergence of the
series --- even for $M_{\rm SUSY} = M_{\tilde{q}}$.
It is interesting to mention that large cancellations
occur among the cubic, quadratic, linear and non-logarithmic term 
of $\hat\Sigma_{\phi_2}$ at three-loop order.
E.g., for our default input values the sum of the cubic and quadratic
logarithm is negative whereas the complete answer leads to a positive
correction for the $\alpha_s^2$ coefficient of $\hat\Sigma_{\phi_2}$.

If we express the result of Eq.~(\ref{eq::sigphi2})
in terms of \drbar{}$^\prime$ parameters, we can compare
with the results obtained in Ref.~\cite{Martin:2007pg}. We find
agreement both for the cubic and quadratic logarithm.



\section{III. Numerical results}


In the remainder of this letter, we discuss the numerical effect of our
result, restricting ourselves to $A_t=0$. We adopt the on-shell scheme
for the quark, squark and gluino masses.

We choose $\mu=M_t$ as the default value for the renormalization scale.
First we compute $\alpha_s(M_t)$, defined in the $\overline{\rm DR}$
scheme and the full \susy{} theory, from the SM input
value $\alpha_s(M_Z)=0.1189$~\cite{Bethke:2006ac} which is given within
five-flavour \qcd{}. We follow the procedure outlined in
Ref.~\cite{Harlander:2007wh} which includes three-loop running and
two-loop matching effects.  As a result we obtain, e.g.,
$\alpha_s(M_t)=0.0926$ for a common SUSY mass $M_{\rm SUSY}=1$~TeV.
The {\abbrev SM} input parameters are given as
$G_F=1.16637\times10^{-5}~\mbox{GeV}^{-2}\,,$ $M_Z =
91.1876~\mbox{GeV}\,$~\cite{Yao:2006px}, $M_t =
170.9~\mbox{GeV}\,$~\cite{:2007bxa}. For the heavy squark mass ($\tilde
q\neq \tilde t$) we use $M_{\tilde{q}}=2$~TeV.

In order to evaluate the tree-level approximation of the Higgs boson
mass we also need the parameters $M_A$ and $\tan\beta$. If not
stated otherwise we adopt the values
$M_A=1~\mbox{TeV}$ and $\tan\beta=40\,.$
Since these parameters do not enter the corrections
considered in this paper, they only have minor
influence on the plots
presented in the following.

\begin{figure}
  \centering
  \begin{tabular}{c}
    \epsfig{file=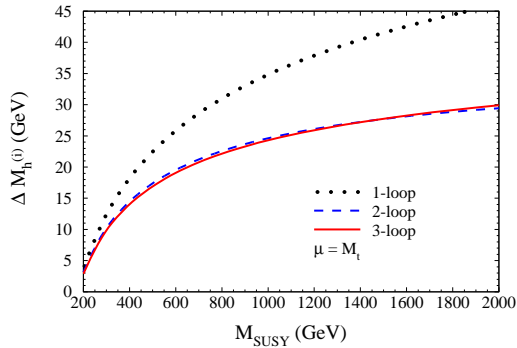,width=.4\textwidth}
  \end{tabular}
  \caption{$\Delta M_h$ as a function of $M_{\rm SUSY}$ at one-, two-,
    and three-loop level. The renormalization scale is set to
    $\mu=\mtop$.  } 
  \label{fig::mhmsusy}
\end{figure}

\begin{figure}
  \centering
  \epsfig{file=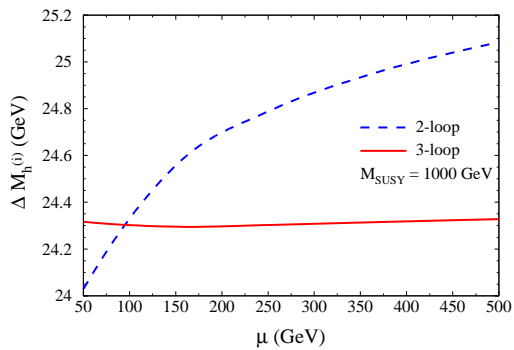,width=.4\textwidth}
  \caption{$\Delta M_h$ as a function of the renormalization scale $\mu$
    at two- and three-loop level, where $M_{\rm SUSY}=1$~TeV has been
    chosen.}
  \label{fig::mhmu}
\end{figure}

In Figs.~\ref{fig::mhmsusy} and \ref{fig::mhmu} we discuss 
the difference between the Higgs boson mass evaluated with $i$-loop
approximation and the tree-level result,
\begin{eqnarray}
  \Delta M_h^{(i)} &=& M_h^{(i-{\rm loop})} - M_h^{\rm tree}
  \,.
\end{eqnarray}
Fig.~\ref{fig::mhmsusy} shows $\Delta M_h^{(i)}$ for $i=1$ (dotted),
$i=2$ (dashed) and $i=3$ (solid line) as a function of $M_{\rm SUSY}$ in
the range between 200~GeV and 2~TeV. As is well known, the one-loop
corrections are large, increasing $M_h$ by up to $46$~GeV. The two-loop
effects are negative, reducing the size of the overall corrections by
about 30\% with respect to the one-loop result. 

The three-loop terms are much smaller and clearly stabilize the
perturbative behaviour.  At $\mu=M_t$, for example, they lead to a
further reduction of $\Delta M_h^{(i)}$ by about $400$~MeV for $M_{\rm
SUSY}=300$~GeV and an enhancement of about 500~\mbox{MeV} for $M_{\rm
SUSY}=2$~TeV.  Note that the numerical impact is larger than the
precision on the lightest Higgs boson mass as expected at the {\abbrev
LHC}.

In order to estimate the size of the higher order corrections, we
consider the dependence of the result on the choice of the
renormalization scale. In Fig.~\ref{fig::mhmu} we plot $\Delta
M_h^{(i)}$ as a function of $\mu$ which is varied from $50$\,GeV to
$500$\,GeV. The two-loop results show a variation of more than 1\,GeV
over this range. The error band derived in this way nicely covers the
three-loop result, which itself varies by less than 35\,MeV. For other
values of $M_{\rm SUSY}$ the variation can reach up to 100~MeV. The
three-loop curve in Fig.~\ref{fig::mhmu} shows a shallow minimum close
to $\mu=M_t$ which in turn is close to the intersection point of the
two- and three-loop result.
This justifies the choice $\mu=M_t$ as default value.


\section{IV. Conclusions}


To summarize, in this letter the three-loop corrections to the lightest
Higgs boson mass have been computed in three different limits of the
\susy{} parameter space. For the phenomenologically interesting case
where the gluino and top squarks have about the same mass and the
remaining squarks are heavier, we observe effects of approximately
500~MeV. The dependence of the three-loop result on the
renormalization scale indicates that the residual theoretical
uncertainty matches the expected accuracy for a Higgs mass measurement
at \lhc{} and possibly even at a future linear collider.

It remains to say that the calculational setup which was used to obtain
the results of this paper is not restricted to the specific \mssm{}
parameter points considered here. A more comprehensive study is in
preparation and will be presented elsewhere.


\noindent
{\bf Acknowledgements.}
We thank Sven Heinemeyer for carefully reading the manuscript and
useful comments and Stephen Martin and Pietro Slavich for
useful communication.
This work was supported by the DFG through SFB/TR~9 and HA 2990/3-1.
We thank the Galileo Galilei Institute for Theoretical Physics for the
hospitality and the INFN for partial support during the completion of
this work.



\end{document}